\documentclass[a4paper]{jpconf}
\usepackage{graphicx}
\usepackage{cite}

\usepackage{bm}

\usepackage{hyperref}
\usepackage{color}

\begin{document}
\title{Hydrodynamic flow in small systems\\[1ex]
{\large{or:  ``How the heck is it possible that a system emitting only\\[-1ex]
\hspace*{11mm} a dozen particles can be described by fluid dynamics?'' }}}

\author{Ulrich Heinz\footnote{This work was supported in part by the U.S. Department of Energy (DOE), Office of Science, Office for Nuclear Physics, under Awards No. \rm{DE-SC0004286}, \rm{DE-FG02-05ER41367}, and through the Beam Energy Scan Theory (BEST) Collaboration, as well as by the National Science Foundation (NSF) within the framework of the JETSCAPE Collaboration under Award No.~ACI-1550223. This work used the Extreme Science and Engineering Discovery Environment (XSEDE), which is supported by National Science Foundation grant number ACI-1548562. Travel support through a bilateral grant from FAPESP and The Ohio State University, as well as the hospitality of the Centre for Theoretical and Mathematical Physics (CTMP) at the University of Cape Town, are gratefully acknowledged.}$^a$, in collaboration with J. Scott Moreland$^b$}
\address{$^a$Department of Physics, The Ohio State University, Columbus, OH 43210-1117, USA\\
$^b$Department of Physics, Duke University, Durham, NC 27708-0305, USA}
\ead{heinz.9@osu.edu}

\begin{abstract}
The ``unreasonable effectiveness'' of relativistic fluid dynamics in describing high energy heavy-ion and even proton-proton collisions will be demonstrated and discussed. Several recent ideas of optimizing relativistic fluid dynamics for the specific challenges posed by such collisions will be presented, and some thoughts will be offered why the framework works better than originally expected. I will also address the unresolved question where exactly hydrodynamics breaks down, and why.
\end{abstract}

\section{Prologue}

In recent years high-energy nuclear collisions at RHIC and the LHC have revealed strong indications for collective flow with hydrodynamic characteristics even in so-called ``small''\footnote{More about those quotation marks later.} collision systems (p-p, p-Au, d-Au, $^3$He-Au, and p-Pb; see e.g. the reviews \cite{Dusling:2015gta, Li:2017qvf, Floris:2019klr}). The question in the subtitle above is one that I get frequently asked in this context. Let me start by explaining that it is the wrong question to ask. To illustrate my point allow me to consider a world without quarks where the strong interaction is described by an SU(3) gauge theory (``QCD'') which contains only gluons in its color-deconfined ``gluon plasma'' state and only glueballs (G) in its color-confined hadronic phase. In such a world a GG collision at LHC energies would create a gluon plasma with similar initial energy ($e$), entropy ($s$) and (if it allows for a quasiparticle description) total particle density ($n$) as the quark-gluon plasma created in a pp collision in our world at the real LHC. The equation of state (EoS) $p(e)$, speed of sound $c_s(e)$, and transport coefficients (such as the specific shear and bulk viscosities $\eta/s$, $\zeta/s$) of this gluon plasma will be very similar to those of the quark-gluon plasma in our world where these quantities are all dominated by the interactions with and among gluons. So the dynamical evolution of the gluon plasma created in GG collisions in this imaginary world will look qualitatively similar to that of the quark-gluon plasma created in pp collisions at the real LHC. If the total entropy of the collision fireball is enough to create, say, a dozen charged hadrons per unit pseudorapidity, corresponding to $dN/d\eta\simeq20$ if you include neutrals, in our real world where most of these hadrons are pions with a rest mass of 140\,MeV, it would not even suffice to create two glueballs G per unit rapidity, $dN_G/d\eta\simeq2$, in the imaginary world if the lightest glueballs had a mass of 1.5\,GeV or more. 

Does this imply that the gluon plasma created in a GG collision with $dN_G/d\eta{\,=\,}2$ in the imaginary world evolves less hydrodynamically than the quark-gluon plasma in a pp collision with $dN_\mathrm{ch}/d\eta{\,=\,}12$ in the real world? Obviously not. That the former collision has much fewer particles in the final state than the latter is a cruel joke of Nature who forces the partition of the system's energy into a small number of very heavy final state hadrons in the glueball world while creating, under the same initial conditions, an order of magnitude more final-state hadrons in our real world. If pions where lighter (say 10\,MeV instead of 140\,MeV), that same pp collision would create about 300 hadrons per unit rapidity in its final state, the same order of magnitude as measured in off-center PbPb collisions at the LHC where few physicists doubt the validity of the hydrodynamic flow paradigm \cite{Heinz:2009xj, Heinz:2013th, Gale:2013da, Romatschke:2017ejr}. The quantization of emitted energy in heavy chunks implies that the underlying fluid dynamical behavior cannot be sampled continuously and suffers from finite number statistical fluctuations --- even more so in the glueball world than in ours --- such that its exploration requires averaging over many similar collision events (same collision system, centrality and collision energy) in order to sample the underlying physics with sufficient statistical precision. So, while the gluon plasma created in the GG collision of our imaginary glueball world may exhibit almost identical hydrodynamic flow patterns to the quark-gluon plasma in a pp collision at the LHC with the same initial entropy per unit rapidity, these patterns would be much harder to discern in the GG collision, due to much larger finite number statistical fluctuations in the final state. This doesn't mean, however, that no such patterns exists -- it is just difficult to distill them from the strongly fluctuating observables.  

I hope that this {\it Gedankenexperiment} convinces you that the absolute value of the number of final state hadrons per unit rapidity is a poor criterium for (pre-)judging the applicability of the hydrodynamic flow paradigm. Final state hadrons are only created at the end of the collision when the quark-gluon plasma hadronizes. Afterwards the hydrodynamic model quickly breaks down, due to the short-range nature of the residual ``strong'' interactions between color-neutral hadrons. The final-state hadrons are not responsible for the interactions that control the system's evolution towards local thermal  equilibrium in its color-deconfined liquid stage -- its EoS, speed of sound, and its transport properties. In its liquid state, the strong open-color interactions in the quark-gluon plasma may even largely invalidate its description in terms of well-defined quasiparticles, again making the number of particle degrees of freedom per unit rapidity a poor criterium for (pre-)judging its ability to develop hydrodynamic flow. The absolute value of  the (initial) entropy per unit space-time rapidity, $dS/d\eta_s= \tau_0 \int d^2r_\perp s(\bm{r}_\perp,\eta,\tau_0)$, on the other hand, which is (on average) monotonically related to the final state charged hadron pseudorapidity density $dN_\mathrm{ch}/d\eta$ \cite{Shen:2015qta}, remains well-defined even in strongly-coupled quantum field theories without good quasi-particles, and thus may be a better starting point for a breakdown criterium of the hydrodynamic paradigm (see, e.g., \cite{Basar:2013hea, Romatschke:2017ejr, Kurkela:2018wud}).

\section{The ``unreasonable effectiveness'' of hydrodynamics for nuclear collisions}

\begin{figure}[h]
\includegraphics[width=\linewidth]{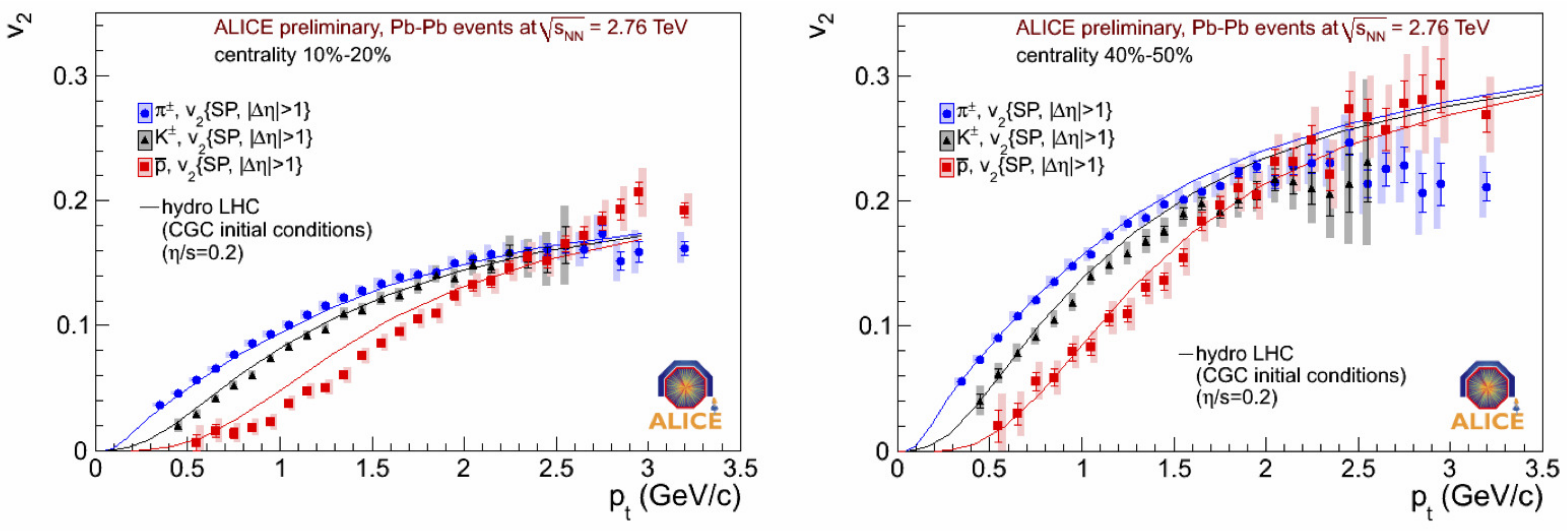}
\caption{\small Differential elliptic flow for pions, kaons and protons in semi-central (left) and semi-peripheral (right) Pb-Pb collisions at the LHC, as reported by the ALICE Collaboration at the Quark Matter 2011 conference \cite{Krzewicki:2011ee}. Solid lines show hydrodynamic \textbf{\textit{pre}}dictions from  \cite{Shen:2011eg}.
\label{F1} }
\end{figure}

\begin{figure}[h]
\includegraphics[width=\linewidth]{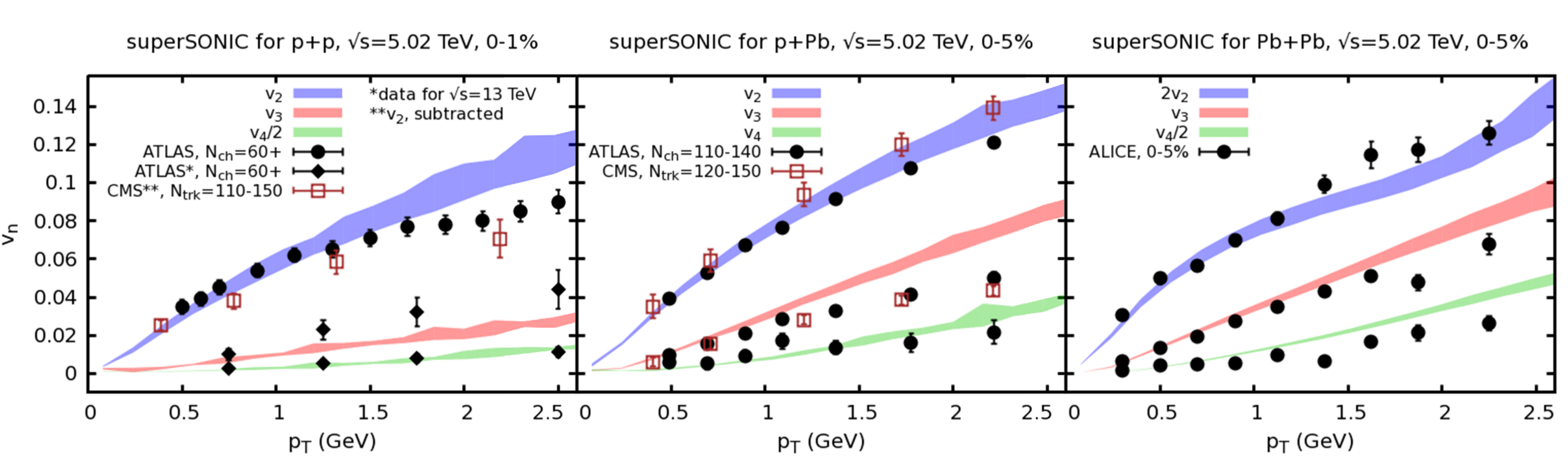}
\caption{\small Differential elliptic ($v_2$), triangular ($v_3$), and quadrangular flow ($v_4$) for charged hadrons from p-p (left), p-Pb (middle) and PbPb (right) collisions at the LHC, compared with hydrodynamic model simulations using the superSONIC code package \cite{Romatschke:2015gxa}. Figure taken from \cite{Weller:2017tsr}. 
\label{F2} }
\end{figure}

In the last decade, relativistic dissipative (``viscous'') fluid dynamics has become the workhorse of dynamical modeling of ultra-relativistic heavy-ion collisions \cite{Heinz:2009xj, Heinz:2013th, Gale:2013da, Romatschke:2017ejr}. In spite of the extraordinarily rapid expansion of the collision fireball, with dramatically different expansion rates along the beam direction (due to the inability of the two colliding nuclei to stop each other \cite{Bjorken:1982qr}) and in the transverse directions where the expansion is driven by pressure gradients and starts from zero, which generates large shear stresses, the hydrodynamic model has proven to possess high predictive power. (An early example is shown in Fig.~\ref{F1}.)
%
%
It works even in ``small'' collision systems, such as p-Pb and p-p at the LHC \cite{Weller:2017tsr} (see Fig.~\ref{F2})
or p-Au, d-Au, and $^3$He-Au at RHIC \cite{PHENIX:2018lia}, as long as subnucleonic fluctuations in the initial energy deposition are appropriately accounted for \cite{Welsh:2016siu}.
%
%
The largest uncertainties in comparing data from such small systems with fluid dynamical code packages (e.g. iEBE-VISHNU \cite{Shen:2014vra} (\textcolor{blue}{\url{https://u.osu.edu/vishnu}}), superSONIC \cite{Romatschke:2015gxa} (\textcolor{blue}{\url{https://sites.google.com/site/revihy/download}}), or MUSIC \cite{Schenke:2010rr} (\textcolor{blue}{\url{http://www.physics.mcgill.ca/music/}})) does not appear to arise from the applicability of the hydrodynamic model, but from our lack of precise knowledge of the internal structure of the nucleon, i.e. the distribution and event-by-event fluctuations of the gluon density inside protons and neutrons \cite{Welsh:2016siu}. 

\begin{figure}[h]
\includegraphics[width=0.72\linewidth]{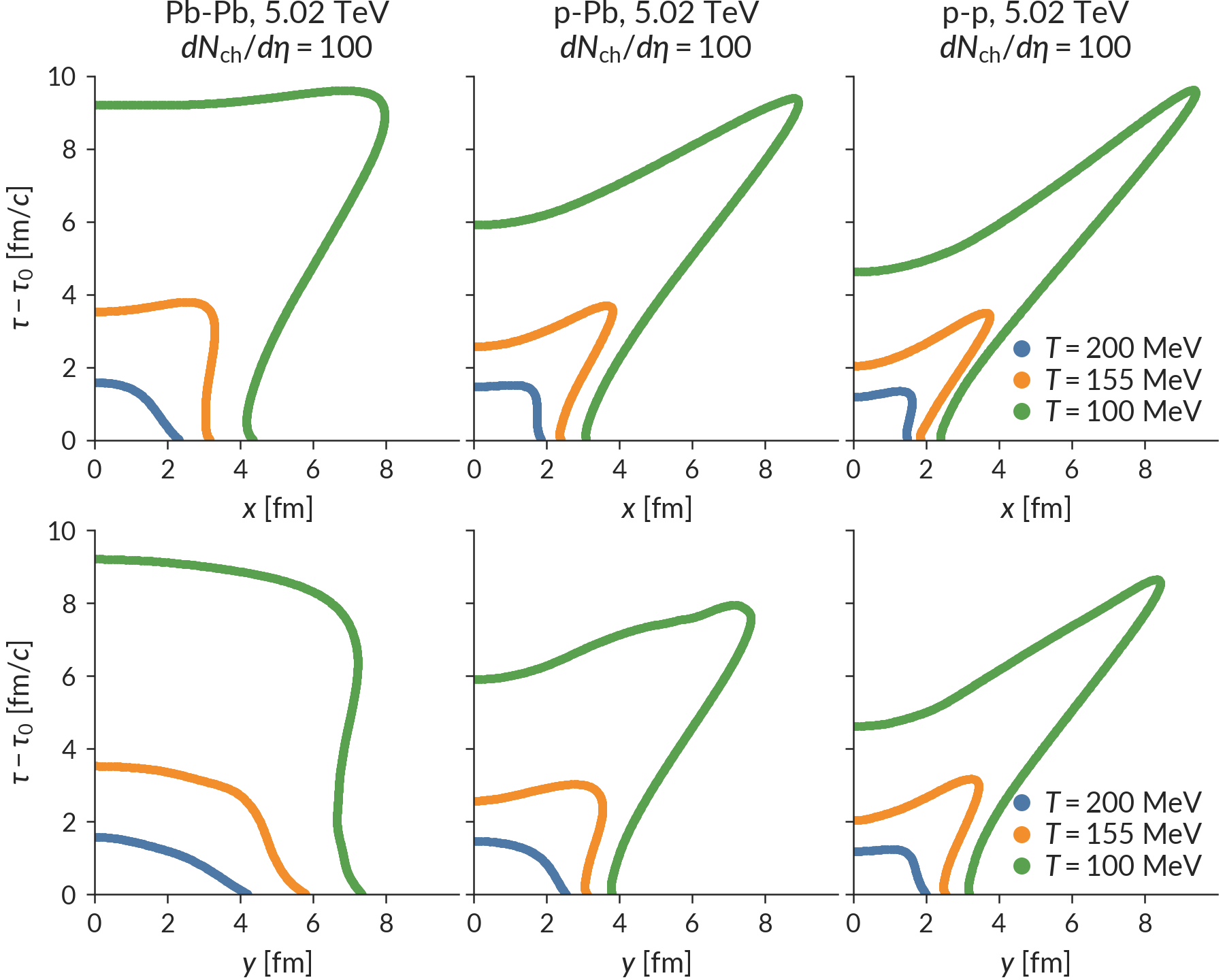}
\includegraphics[width=0.265\linewidth]{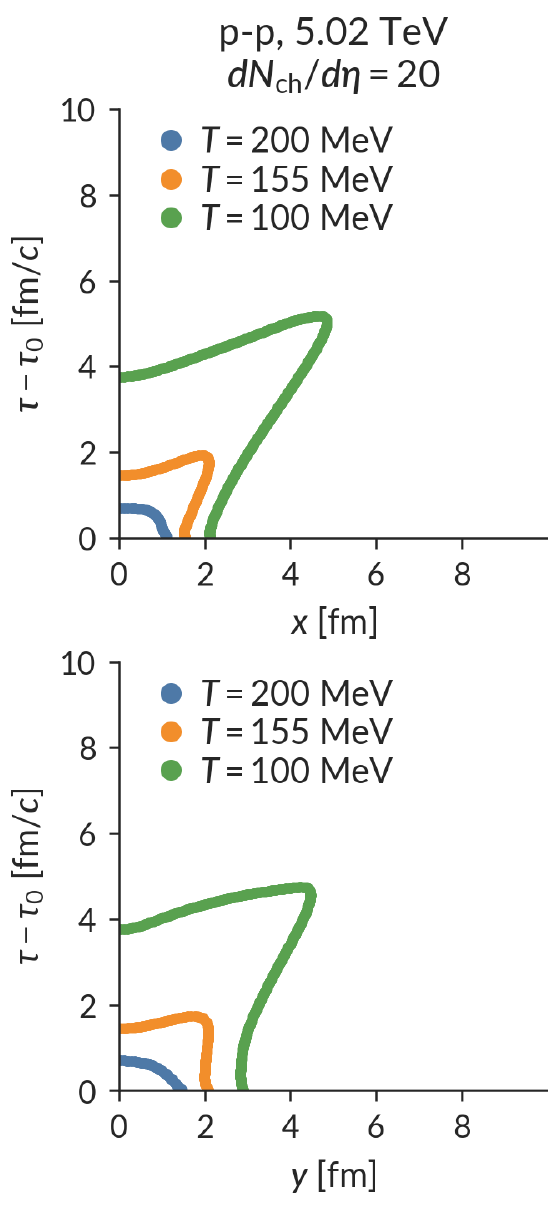}
\caption{\small Isotherms of temperatures $T=200$ (blue), 155 (orange) and 100\,MeV (green) along the short ($x$, top row) and long ($y$, bottom row) directions of an ensemble-averaged fireball constructed from elliptically deformed and aligned fluctuating initial entropy density profiles from the T$_\mathrm{R}$ENTo model with exponent $p=0$ \cite{Moreland:2014oya} (similar to IP-GLASMA initial conditions) for Pb-Pb (left), p-Pb (center-left), and p-p collisions (center-right and right columns) at $\sqrt{s_{_\mathrm{NN}}}=5.02$\,TeV. The three left columns compare events from the three different collision systems with the same charged hadron pseudorapidity density $dN_\mathrm{ch}/d\eta=100$ in the final state. The right column shows for comparison the corresponding isotherms for p-p collisions at the same collision energy but with five times smaller final pseudorapidity density $dN_\mathrm{ch}/d\eta=20$. All events were evolved with iEBE-VISHNU \cite{Shen:2014vra} using transport coefficients and other model parameters determined by Bayesian model calibration \cite{Bernhard:2016tnd}.
\label{F3}}
\end{figure}

Let me explain now why I use quotation marks when writing about ``small'' collision systems. As argued above, a good starting point for (pre-)judging the applicability of hydrodynamics is the total entropy per unit space-time rapidity $dS/d\eta_s$ deposited in the collision zone. Although, for a given collision configuration, $dS/d\eta_s$ is monotonically related to the final charged hadron pseudorapidity density $dN_\mathrm{ch}/d\eta$, the proportionality constant depends on the additional entropy produced by viscous heating during the expansion, and the latter increases with the fireball expansion rate. Fig.~\ref{F3} compares isotherms along the short and long directions of elliptically deformed fireballs created in peripheral Pb-Pb, central p-Pb, and high-multiplicity p-p collisions at $\sqrt{s_{_\mathrm{NN}}}=5.02$\,TeV with the same final charged hadron pseudorapidity density $dN_\mathrm{ch}/d\eta=100$. (For comparison the right column shows the corresponding isotherms for less extreme p-p collisions with a five times smaller final multiplicity.)    
%
%
This comparison illustrates a number of important points: First, the fireballs created in the so-called ``small'' collision systems p-Pb and p-p are only initially small, due to the small cross section of the proton. As long as roughly the same total entropy $dS/d\eta_s$ is deposited initially, they all have roughly the same (much larger) size at hadronization and at  freeze-out. This is easy to understand: since constant temperature implies constant density, identical multiplicities must correspond to identical volumes. Therefore, events with the same final multiplicity $dN_\mathrm{ch}/d\eta$ have the same freeze-out volume, irrespective of how dilute or compact the fireball's initial configuration was.  

Second, however, if the initial entropy $dS/d\eta_s$ is initially deposited within a smaller transverse area, the larger transverse pressure gradients in this more compact initial configuration drives stronger radial transverse flow \cite{Hirono:2014dda,Kalaydzhyan:2015xba}, reflected in the larger rate of growth of the outer radius of the isotherms in the center-left and center-right columns of Fig.~\ref{F3} compared to the left column. The resulting higher expansion rate reduces the ``Hubble-volume'' of the expanding fireball whose dimensions (``HBT radii'') can be measured with two-particle intensity interferometry \cite{Heinz:1996rw}. At the same final multiplicity and volume, therefore, p-p collisions feature smaller HBT radii than p-Pb collisions, and p-Pb collisions have smaller HBT radii than Pb-Pb collisions. This effect has been observed and noted by the ALICE Collaboration (see Fig.~9 in \cite{Adam:2015pya}).

A popular criterium for the validity of fluid dynamics is the Knudsen number, defined as the ratio between the microscopic interaction length (``mean free path'') and the macroscopic hydrodynamic length scale. In expanding systems, this macroscopic length scale is not given by the total radius of the fireball (related to the total freeze-out volume) but by its Hubble radius (which can be expressed through the expansion rate or through appropriately normalized space-time gradients of the energy or entropy density). This suggests that any phenomenological criterium for applicability of hydrodynamics should involve, in addition to the observed charged hadron multiplicity $dN_\mathrm{ch}/d\eta$ (as a proxy for the initial entropy density $dS/d\eta_s$), the HBT radii or, better, the cube root of the HBT volume $\left |\det(R_{ij}^2)\right |^{1/6}$ (as a proxy for the Hubble radius).  

As an aside I note that the higher expansion rate causes stronger viscous heating in the ``small'' collision systems. So, if the three systems shown in Fig.~\ref{F3} had been initialized with the same initial entropy per unit rapidity $dS/d\eta_s$, the final entropy (and thus $dN_\mathrm{ch}/d\eta$) and final total volumes would be somewhat larger for p-Pb than for Pb-Pb, and still larger for p-p collisions. This would have further exacerbated the above-mentioned effects on the radial flow and HBT radii. 

\begin{figure}[h]
\includegraphics[width=0.33\linewidth]{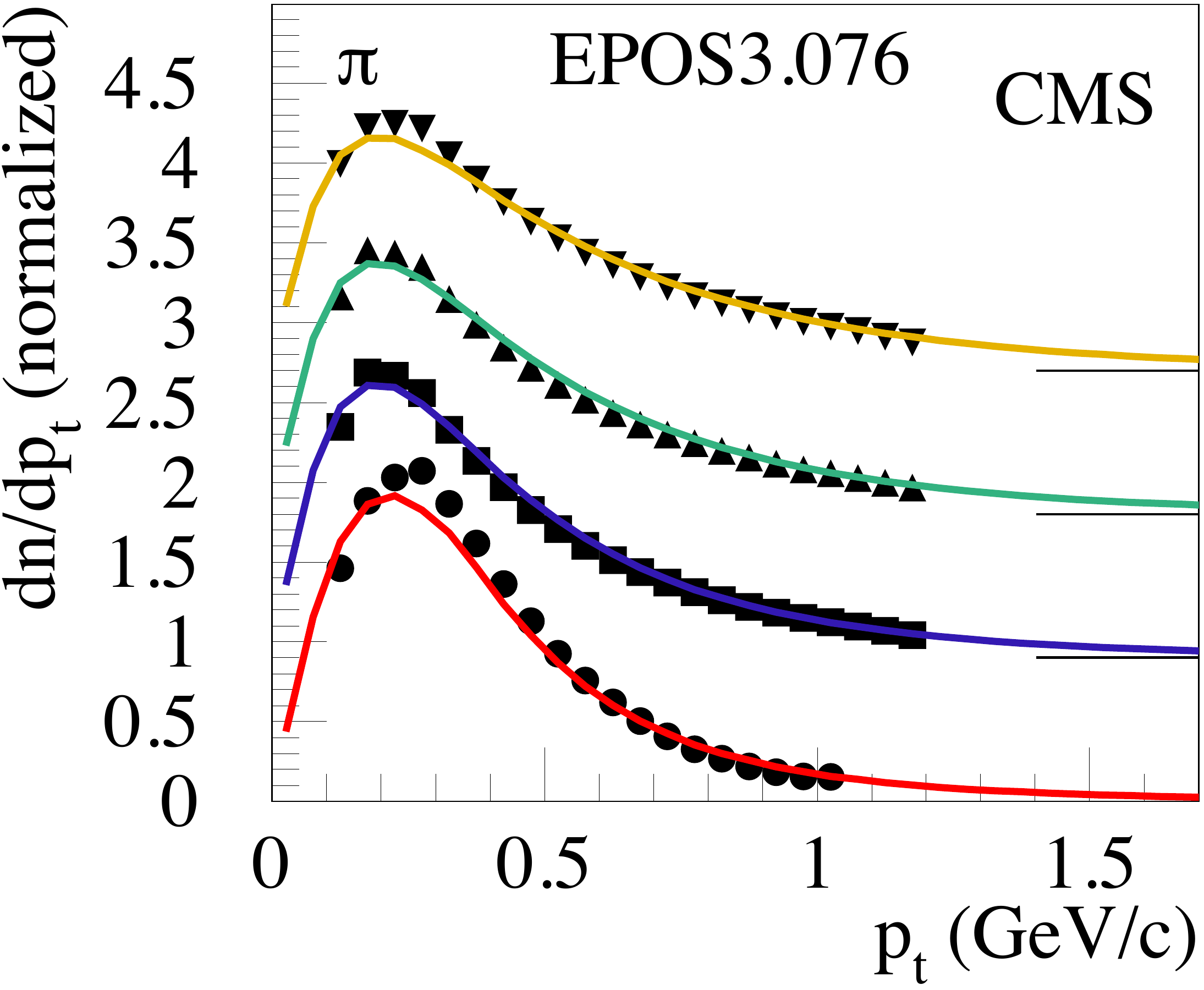}
\includegraphics[width=0.33\linewidth]{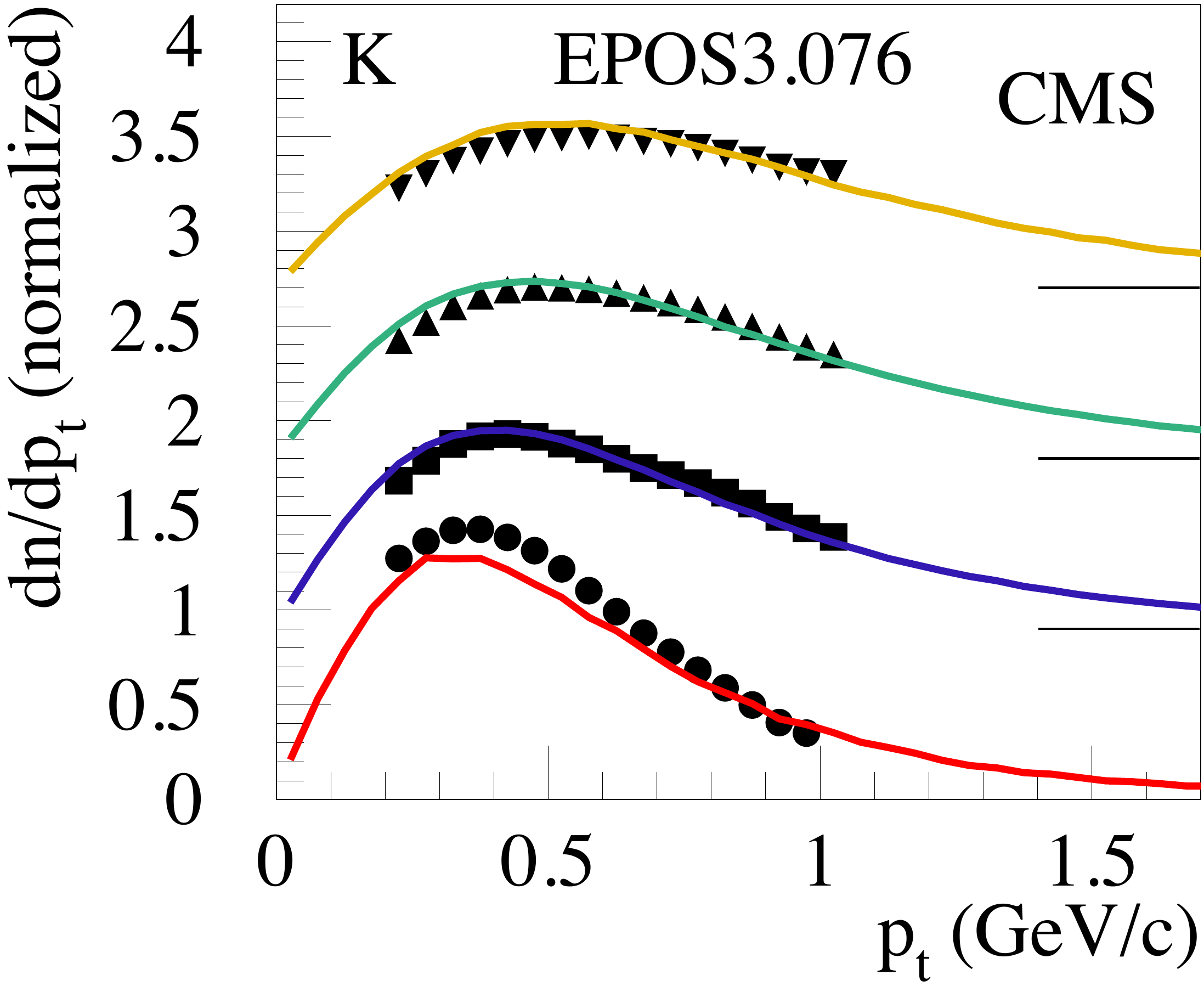}
\includegraphics[width=0.33\linewidth]{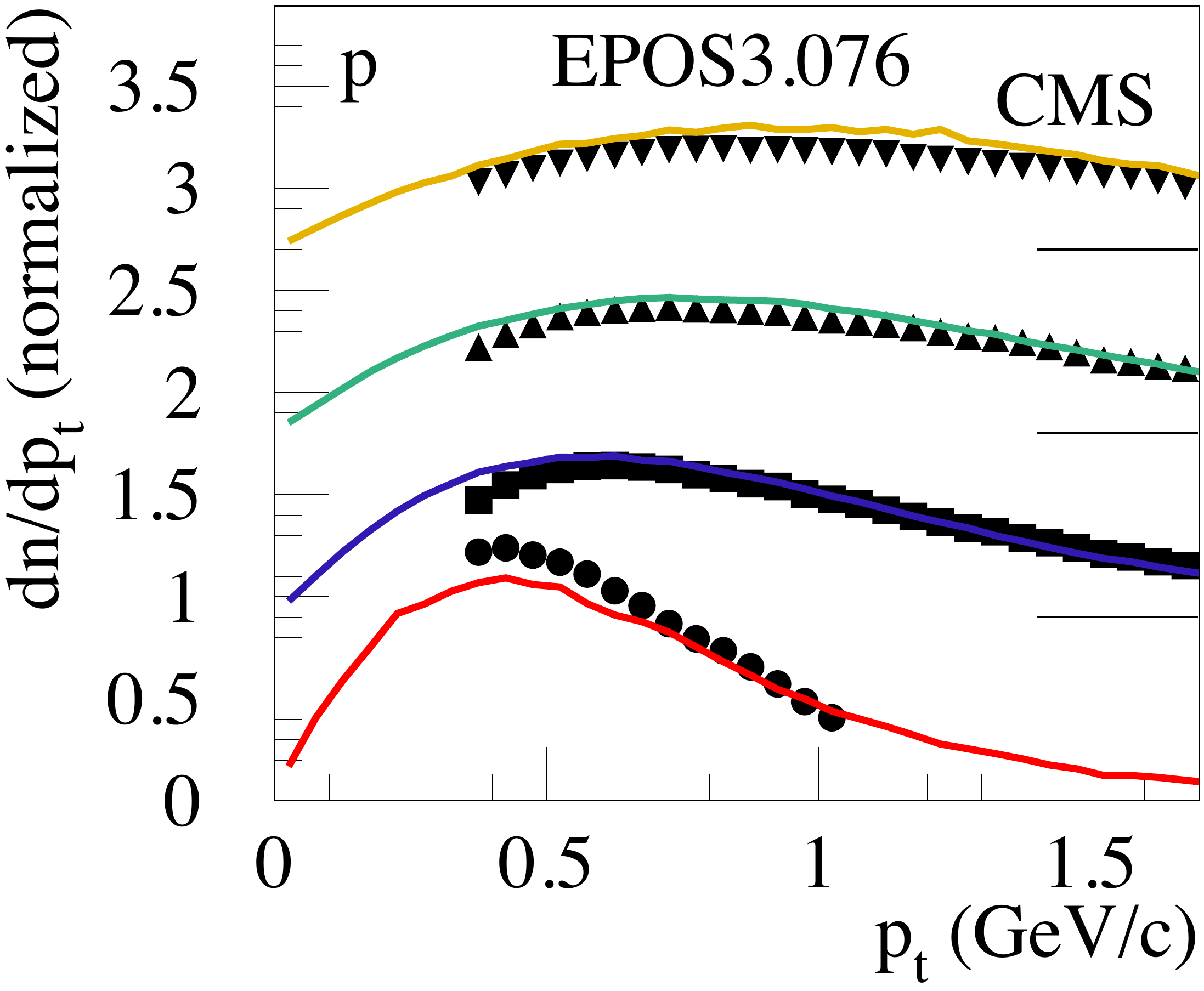}
\caption{\small Transverse momentum spectra of pions, kaons and protons from p-Pb collisions at 5.02\,TeV with different numbers of charged tracks in $|\eta|<2.4$ (8 (dots), 84 (squares), 160 (triangles), 235 (inverted triangles)) as measured by CMS \cite{Chatrchyan:2013eya}. The lines show numerical simulations with the EPOS3.076 code which features a viscous fluid dynamic core \cite{Werner:2013tya}. Similar behavior was found in by CMS in p-p collisions at 0.9, 2.76 and 7\,TeV \cite{Chatrchyan:2012qb}. Figure taken from \cite{Werner:2013tya}.
\label{F4} }
\end{figure}

That initially denser collision systems develop stronger radial flow, as predicted by hydrodynamics, can be directly seen in Fig.~\ref{F4}. 
%
%
It shows pion, kaon and proton spectra from 5.02\,TeV p-Pb collisions measured by CMS for four multiplicity bins. As the multiplicity increases, the spectra become flatter (``harder''), and the effect increases with particle rest mass. This is a hallmark of radial flow \cite{Lee:1990sk,Schnedermann:1993ws} which (i) increases with the initial entropy density and (ii) pushes hadrons out to larger $p_T$ values by an amount that increases with their rest mass. For the same final multiplicity and collision energy, the effect is stronger in p-p collisions \cite{Chatrchyan:2012qb} than in p-Pb collisions \cite{Chatrchyan:2013eya}. The measured behavior can be quantitatively described by EPOS3.076 which has a viscous hydrodynamic core. That same model also describes the elliptic flow (``double ridge'') discovered by CMS in high-multiplicity p-p collisions at 7\,TeV \cite{Khachatryan:2010gv} and later confirmed by both ATLAS \cite{Aad:2015gqa} and CMS \cite{Khachatryan:2015lva} in p-p collisions at 13\,TeV, whose collective nature was demonstrated by CMS by measuring it with 4- and 6-particle cumulants \cite{Khachatryan:2016txc}. 
  
\section{Far-from-equilibrium hydrodynamics}

The discussion above has established that (i) viscous fluid dynamics is phenomenologically very successful and yields quantitatively precise descriptions of and predictions for soft-hadron spectra and flow correlations in Au-Au at RHIC and Pb-Pb at the LHC while providing at least a semiquantitative description of the same observables in p-Au, d-Au, $^3$He-Au, p-Pb and even high-multiplicity p-p collisions, while at the same time (ii) being characterized by large dissipative effects caused by the extremely rapid and anisotropic expansion of the heavy-ion collision fireballs. In fact, the approach is now being successfully used for extracting, with quantified uncertainties, key parameters characterizing the thermodynamic and transport properties of quark-gluon plasma from a global model-to-data comparison with advanced Bayesian statistical analysis tools \cite{Bernhard:2016tnd,Bernhard:2018hnz}. Why do these large dissipative corrections not destroy the precision and predictive power of the hydrodynamic approach?

In the last part of my presentation I will cover some work that my collaborators and I have performed over the last few years to address the particular challenges faced by hydrodynamic approaches when applied to relativistic heavy-ion collisions. These studies uncovered several surprises that showed that the hydrodynamic approach is much more robust and resilient than originally expected. We now understand that its applicability requires neither local thermalization ({\it i.e.}\ thermalized exponential momentum distributions in the local rest frame) nor even local momentum isotropy. This is a dramatic change in our understanding compared to 20 years ago when it was believed (certainly by me!) that the good agreement between ideal fluid dynamics and RHIC data \cite{Ackermann:2000tr} implied very short thermalization times of order $<1$\,fm/$c$ \cite{Heinz:2001xi}. We now understand that this time characterizes the time scale of ``hydrodynamization'' at which the system enters the region of validity of a second-order viscous hydrodynamic approach, rather than real local thermalization at which the fluid would obey the laws of ideal fluid dynamics. In other words, \textbf{\textit{dissipative hydrodynamics works even far from local thermal equilibrium, with quantitative precision}}.

Ultra-relativistic heavy-ion collision dynamics pose two specific challenges to the applicability of dissipative fluid dynamics \cite{Strickland:2014pga, Heinz:2015gka, McNelis:2018jho}: (i) a large shear-viscous stress, in the form of a large difference $P_\perp{-}P_L$ between the transverse and longitudinal pressures, caused by large initial anisotropies between the longitudinal and transverse expansion rates, and (ii) a possibly large bulk viscous pressure $\Pi$ caused by critical dynamics near the quark-hadron phase transition. Optimized hydrodynamic approaches, such as anisotropic hydrodynamics \cite{Martinez:2010sc, Florkowski:2010cf, Bazow:2013ifa, Strickland:2014pga, Tinti:2015xwa, Molnar:2016gwq, McNelis:2018jho} can handle these challenges more efficiently than standard dissipative fluid dynamics.

Hydrodynamics is an effective theory whose form is independent of the strength of the microscopic interactions. Hydrodynamic equations can thus be derived from kinetic theory in a window of weak coupling and small pressure gradients where both approaches are simultaneously valid. 
Only the values of the transport coefficients and the equation of state depend on the microscopic coupling strength;  for the strongly coupled quark-gluon plasma created in heavy-ion collisions, they must be obtained with non-perturbative methods.

In kinetic theory, the conserved macroscopic currents $j^\mu(x)=\langle p^\mu\rangle(x)$  (particle current) and $T^{\mu\nu}(x)=\langle p^\mu p^\nu\rangle(x)$ (energy-momentum tensor) are obtained by taking momentum moments $\langle O(p)\rangle(x)\equiv\frac{g}{(2\pi)^3}\int\frac{d^3p}{E_p}\, O(p) f(x,p)$ of the distribution function $f(x,p)$. Hydrodynamic equations are obtained by splitting the distribution function into a leading-order contribution $f_0$, parametrized through macroscopic observables as
\begin{equation}
\label{eq1}
f_0(x,p)=f_0\left(\frac{\sqrt{p_\mu\Omega^{\mu\nu}(x)p_\nu}-\tilde\mu(x)}{\tilde{T}(x)}\right),
\end{equation}
and a smaller first-order correction $\delta f$ ($|\delta f/f_0|\ll1$):
\begin{equation}
\label{eq2}
f(x,p)=f_0(x,p)+\delta f(x,p).
\end{equation}
In Eq.~(\ref{eq1}), $p_\nu\Omega^{\mu\nu}(x)p_\nu=m^2+\bigl(1+\xi_\perp(x)\bigr) p_{\perp,\mathrm{LRF}}^2 + \bigl(1+\xi_L(x)\bigr) p_{z,\mathrm{LRF}}^2$, where the hydrodynamic flow field $u^\mu(x)$ defines the local fluid rest frame (LRF). $\tilde{T}(x)$ and $\tilde{\mu}(x)$ are the effective temperature and chemical potential in the LRF, Landau matched to the energy and particle densities, $e$ and $n$ \cite{Heinz:2015gka}. $\xi_{\perp,L}$ parametrize the momentum anisotropy in the LRF and are Landau matched to the transverse and longitudinal pressures, $P_T$ and $P_L$ \cite{Tinti:2015xwa, Molnar:2016gwq, McNelis:2018jho}. The latter encode the bulk viscous pressure $\Pi=(2P_\perp{+}P_L)/3-P_\mathrm{eq}$ and the largest shear stress component  $P_\perp{-}P_L$. In anisotropic hydrodynamics, $P_\perp$ and $P_L$ evolve macroscopically according to equations that reflect the competition between macroscopic anisotropic expansion (driving the system away from local equilibrium and momentum isotropy) and microscopic scattering (trying to restore them) \cite{McNelis:2018jho}.

Using the decomposition (\ref{eq2}) we write $T^{\mu\nu}= T^{\mu\nu}_0+\delta T^{\mu\nu}\equiv T^{\mu\nu}_0+\Pi^{\mu\nu}$, $j^\mu= j^\mu_0+\delta j^\mu\equiv j^\mu_0+V^\mu$. Different hydrodynamic approaches can be characterized by the assumptions they make about the dissipative corrections and/or the approximations they use to derive their dynamics from the underlying Boltzmann equation:\\[0.5ex]
{\bf 1.\ Ideal hydrodynamics} assumes local momentum isotropy, setting $f_0$ to be isotropic ($\xi_{\perp,L}=0$) and all dissipative currents to zero: $\Pi^{\mu\nu}=V^\mu=0$.\\[0.5ex]
{\bf 2.\ Navier-Stokes (NS) theory} maintains local momentum isotropy at leading order and postulates instantaneous constituent relations for $\Pi^{\mu\nu}$ and $V^\mu$ by introducing viscosity and heat conduction as transport coefficients that relate these flows to their driving forces. It ignores the microscopic relaxation time that is needed for these flows to adjust to their Navier-Stokes values, leading to acausal signal propagation.\\[0.5ex]
{\bf 3.\ Israel-Stewart (IS) theory} \cite{Israel:1979wp} improves on NS theory by evolving  $\Pi^{\mu\nu}$ and $V^\mu$ dynamically, with evolution equations derived from moments of the Boltzmann equation, keeping only terms linear in the Knudsen number $\mathrm{Kn}=\lambda_\mathrm{mfp}/\lambda_\mathrm{macro}$.\\[0.5ex]
{\bf 4.\ Denicol-Niemi-Molnar-Rischke (DNMR) theory} \cite{Denicol:2012cn} improves IS theory by keeping nonlinear terms up to order $\mathrm{Kn}^2$ and $\mathrm{Kn}\cdot\mathrm{Re}^{-1}$ when evolving $\Pi^{\mu\nu}$ and $V^\mu$.\\[0.5ex]
{\bf 5.\ Third-order Chapman-Enskog expansion} \cite{Jaiswal:2013vta} keeps terms of up to third order when evolving $\Pi^{\mu\nu}$ and $V^\mu$.\\[0.5ex]
{\bf 6.\ Anisotropic hydrodynamics ({\sc aHydro})} \cite{Martinez:2010sc, Florkowski:2010cf} allows for a leading-order local momentum anisotropy ($\xi_{\perp,L}\ne0$), evolved according to equations obtained from low-order moments of the Boltzmann equation, but ignores residual dissipative flows: $\Pi^{\mu\nu}=V^\mu=0$.\\[0.5ex] 
{\bf 7.\ Viscous anisotropic hydrodynamics ({\sc vaHydro})} \cite{Bazow:2013ifa, Bazow:2015cha} improves on {\sc aHydro} by additionally evolving (using IS or DNMR theory) the residual dissipative flows $\Pi^{\mu\nu},\,V^\mu$ generated by the deviation $\delta f$ around the locally anisotropic leading distribution function $f_0$.

There exist a few highly symmetric situations for which the Boltzmann equation, in Relaxation Time Approximation (RTA), can be solved exactly. These include the Bjorken \cite{Bjorken:1982qr} and Gubser \cite{Gubser:2010ze} flows which are (although highly idealized) relevant for heavy-ion collisions \cite{Denicol:2014xca, Denicol:2014tha}. While for Bjorken expansion the expansion rate decreases with longitudinal proper time $\tau$ like $1/\tau$, allowing the system to asymptotically reach local momentum isotropy and thermal equilibrium, Gubser expansion includes an additional strong transverse flow which leads to an asymptotically constant expansion rate and exponentially growing Knuden number Kn \cite{Denicol:2014tha}, and thus to asymptotic free-streaming. The exact evolution of the macroscopic currents $T^{\mu\nu}$ and $j^\mu$ associated with these solutions can be compared with that predicted by any of the 7 different hydrodynamic approximations listed above and thus be used to assess the accuracy of the latter in these two opposite extremes of asymptotic evolution.

To illustrate the differences between the different hydrodynamic approximations, we briefly summarize the corresponding evolution equations for the shear stress. (For both Bjorken and Gubser flow with a conformal equation of state $\Pi^{\mu\nu}$ has only one independent component, the shear stress $\pi^{\eta\eta}$.) For Gubser flow (where all macroscopic quantities depend on only one space-time variable, the de Sitter time $\rho$  \cite{Gubser:2010ze}) one finds the following 
\cite{Marrochio:2013wla, Denicol:2014xca, Denicol:2014tha, Nopoush:2014qba, Martinez:2017ibh, Chattopadhyay:2018apf} (a similar discussion for Bjorken flow can be found in \cite{Florkowski:2013lza, Florkowski:2014sfa, Bazow:2013ifa}):\footnote{All quantities with hats have been made unitless by multiplying with appropriate powers of the proper time $\tau$.}\\[0.5ex]
{\bf 1.\ Ideal hydrodynamics} gives $\hat{T}_{\mathrm{ideal}}(\rho) = \frac{\hat{T}_0}{\cosh^{2/3}(\rho)}$, combined with zero shear stress, $\hat{\pi}^{\eta\eta}{\,=\,}0$.\\[0.5ex]
%
%
\begin{figure}[b!]
\vspace*{-3mm}
\centering
\includegraphics[width=0.91\linewidth]{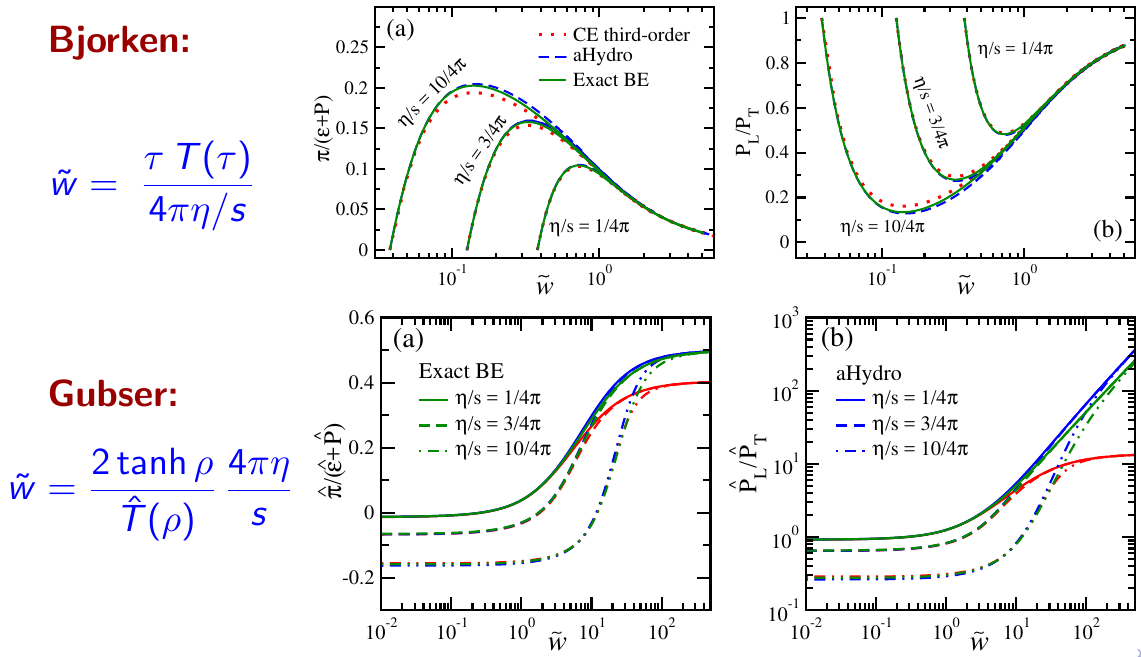}
\vspace*{-3mm}
\caption{\small Normalized shear stress $\pi/(e{+}P_\mathrm{eq})=\frac{1}{4}(\pi/P_\mathrm{eq})$
(left column) and pressure anisotropy $P_L/P_\perp$ (right column) for Bjorken (upper row) and Gubser flow (lower row), plotted as functions of a rescaled time variable $\tilde w$ which for Bjorken flow corresponds to the inverse Knudsen number, $\tilde w{\,=\,}\rm{Kn}^{-1}$, and for Gubser flow directly to the Knudsen number, $\tilde w$\,=\,Kn \cite{Chattopadhyay:2018apf}.  
\label{F5} }
\end{figure}
%
{\bf 2.-7.} For all dissipative hydrodynamic frameworks the temperature evolves instead according to the differential equation $\frac{1}{\hat{T}}\frac{d\hat{T}}{d\rho}+\frac{2}{3}\tanh \rho = \frac{1}{3}\bar{\pi}_{\eta}^{\eta}(\rho )\tanh\rho$ \cite{Gubser:2010ze,Marrochio:2013wla} where $\bar{\pi}\equiv \hat{\pi}_\eta^\eta/(\hat{T}\hat{s})$. Differences between the approaches arise from their different evolution of the shear stress. In\\
{\bf 2.\,NS theory} the shear stress is given by the (instantaneous) constituent relation $\hat{\pi}_{NS}^{\eta\eta}=\frac{4}{15}\hat{\tau}_\mathrm{rel}\tanh \rho$ \cite{Denicol:2014tha} where $\hat{\tau}_\mathrm{rel}=\mathrm{const.}/\hat{T}$. In all second and higher-order hydrodynamic approaches the shear stress evolves instead according to a differential equation of the type 
\begin{equation}
\label{eq3}
  d\bar{\pi}_{\eta}^{\eta}/d\rho + \bar{\pi}_\eta^\eta/\hat{\tau}_\mathrm{rel} = 
  \bigl(a_1 + a_2 \bar{\pi} - a_3 \bar{\pi}^{2}\bigr)\tanh \rho - \frac{4}{3} {\cal F}(\bar{\pi}).
\end{equation}
For the approaches {\bf 3., 4., 5.} ({\it i.e.} as long as the equations are derived by expanding around a locally isotropic distribution function $\xi_{\perp,L}{\,=\,}0$) the function ${\cal F}$ vanishes: ${\cal F}(\bar{\pi})=0$. Only for {\bf 7.} {\sc vaHydro} ${\cal F}(\bar{\pi})$ is nonzero; its specific form is found in \cite{Martinez:2017ibh}.\footnote{The {\sc aHydro} study in \cite{Nopoush:2014qba} expresses the evolution of the shear stress in terms of the microscopic parameters $\xi_{\perp,L}$ and thus cannot be directly compared to the macroscopic evolution equation (\ref{eq3}).} For the constants $(a_1,a_2,a_3)$ one finds:\\[0.5ex]
{\bf 3.\ IS theory:} $(a_1,a_2,a_3) = \left(\frac{4}{15},\, 0,\,\frac{4}{3}\right)$.\\
{\bf 4.\ DNMR theory:} $(a_1,a_2,a_3) = \left(\frac{4}{15},\, \frac{10}{21},\,\frac{4}{3}\right)$.\\
{\bf 5.\ Third-order Chapman-Enskog expansion:} $(a_1,a_2,a_3) = \left(\frac{4}{15},\, \frac{10}{21},\,\frac{412}{147}\right)$ \cite{Chattopadhyay:2018apf}.\\
{\bf 6.\ Anisotropic hydrodynamics ({\sc aHydro}):} See footnote 4.\\
{\bf 7.\ Viscous anisotropic hydrodynamics {\sc vaHydro}:} $(a_1,a_2,a_3) = \left(\frac{5}{12},\, \frac{4}{3},\,\frac{4}{3}\right)$ \cite{Martinez:2017ibh}.

Figure~\ref{F5} shows, for thermal equilibrium initial conditions, the time evolution of the normalized shear stress $\bar\pi$ and the pressure anisotropy $P_L/P_\perp$ for Bjorken and Gubser flows, for three systems with specific shear viscosities $4\pi\eta/s=4\pi T\tau_\mathrm{rel}/5=1,\,3$, and 10 \cite{Chattopadhyay:2018apf}. For clarity, the exact solution of the Boltzmann equation (green solid lines) is compared only with the two best-performing hydrodynamic approximations, the third-order Chapman-Enskog expansion (red dotted lines) and anisotropic hydrodynamics (blue dashed lines) [where in this case, due to the high degree of symmetry of the flow, {\sc aHydro} and {\sc vaHydro} correspond to the same approximation \cite{Martinez:2017ibh}]. Similar comparisons for the other hydrodynamic approximations discussed in this contribution can be found in the literature \cite{Marrochio:2013wla, Florkowski:2013lya, Florkowski:2013lza, Bazow:2013ifa, Denicol:2014xca, Denicol:2014tha, Nopoush:2014qba, Florkowski:2014sfa, Strickland:2014pga, Heller:2016rtz, Martinez:2017ibh, Romatschke:2017vte, Alqahtani:2017mhy, Strickland:2017kux, Chattopadhyay:2018apf}. Following \cite{Heller:2016rtz} the time variables $\tau$ (for Bjorken flow) and $\rho$ (for Gubser flow) are replaced by a scaling variable $\tilde{w}$, defined as the product of the macroscopic expansion rate (${=\,}1/\tau$ in the case of Bjorken flow and ${=\,}2\tanh\rho$ for Gubser flow) with the microscopic relaxation time $\tau_\mathrm{rel}{\,=\,}4\pi\eta/(Ts)$ (Gubser flow), or as its inverse (Bjorken flow). The idea behind this rescaling is that for Bjorken flow the system approaches thermalization ({\it i.e.}\ a regime of small Knudsen numbers) at late times while for Gubser flow it becomes asymptotically free-streaming ({\it i.e.}\ approaches a regime of large Knudsen numbers at late times). The rate of this approach scales with the microscopic relaxation time. 

Figure~\ref{F5} shows that for Bjorken flow (top row) the solutions of the Boltzmann equation and of the two hydrodynamic approximations shown in the plot approach a common attractor \cite{Heller:2015dha} at late times where the system approaches local momentum isotropy and thermal equilibrium. Different initial conditions relax exponentially towards  this attractor. However, the hydrodynamic models describe the exact Boltzmann dynamics well even at early times where, for $\eta/s{\,=\,}10/(4\pi)$, the systems moves very far away from equilibrium, as witnessed by the shear pressure becoming almost as large as the thermal pressure (corresponding to a large inverse Reynolds number Re$^{-1}={\cal O}(1)$). 

In the bottom row of the figure one sees that the approach to a common late-time attractor persists in the case of Gubser flow, at least for anisotropic hydrodynamics while the third-order Chapman-Enskog approach approaches an incorrect asymptotic value for the inverse Reynolds number ($\pi/P_\mathrm{eq}\to1.6$ instead of 2). Still, the third-order Chapman-Enskog approach performs much better than all other hydrodynamic approximation schemes that are based on expansions around local momentum isotropy. That anisotropic hydrodynamics correctly reproduces even the asymptotic free-streaming limit of Gubser flow is a striking counterexample to the folklore that hydrodynamics can only be applied to systems that are close to local momentum isotropy and thermal equilibrium. Still, the high quality of the anisotropic hydrodynamic description of Gubser flow by {\sc aHydro} (blue-dashed lines) may be somewhat accidental in that Gubser symmetry may produce phase-space distributions that are particularly well adjusted to being decomposed as in Eqs.\ (\ref{eq1}) and (\ref{eq2}). Upcoming (3+1)-dimensional studies \cite{McNelis:2018jho} will shed further light on this issue.

\section*{References}

\bibliography{HI}

\end{document}